\documentclass[11pt]{article}
\pdfoutput=1
\usepackage{jheppub}
\usepackage{lmodern}
\usepackage{amsmath,amssymb,mathtools,amsthm}
\usepackage{bm}
\usepackage{booktabs,array}
\usepackage{enumitem}
\usepackage{microtype}
\usepackage{xcolor}
\newcommand{\cS}{\mathcal S}

\newcommand{\cC}{\mathcal C}
\newcommand{\tr}{\operatorname{tr}}

\newcommand{\im}{\operatorname{im}}

\theoremstyle{plain}
\newtheorem{theorem}{Theorem}[section]

\begin{document}
\title{Fortuitous BMN Cohomology at Every $N$}
\author[a,b]{Robert de Mello Koch,}
\author[a,c]{Minkyoo Kim,}
\author[d]{and Sinong Liu}
\affiliation[a]{School of Science, Huzhou Normal University, Huzhou 313000, China}
\affiliation[b]{Mandelstam Institute for Theoretical Physics, School of Physics, University of the Witwatersrand, Private Bag 3, Wits 2050, South Africa}
\affiliation[c]{Research Institute of Basic Sciences, Seoul National University, Seoul 08826, Republic of Korea}
\affiliation[d]{Department of Physics, Kyung Hee University, 26, Kyungheedae-ro, Dongdaemun-gu, Seoul, 02447, Republic of Korea}
\emailAdd{robert@zjhu.edu.cn}
\emailAdd{mimkim80@gmail.com}
\emailAdd{sinong.liu@khu.ac.kr}
\abstract{Finite-$N$ trace relations can produce $Q$-cohomology classes with no stable large-$N$ ancestor. The brane-fused construction at $SU(3)$ suggests that such states can retain an $SU(2)$ nonabelian core accompanied by scalar hair. We develop this idea into a general construction for every $SU(N)$, $N\geq3$. Starting from the $SU(2)$ threshold operator $O_{24}$, we select pairs of eigenvalues of one scalar and construct corrected two-colour letters whose $Q$-transformations obey the $SU(2)$ rules exactly. Substitution into $O_{24}$ gives a closed rational operator for each pair. Multiplying by eigenvalue-separation factors removes the denominators, and summing over all pairs produces a gauge-invariant polynomial. An exact block-restriction identity recovers a nonzero multiple of $O_{24}$. The restriction maps every stable class to a stable $SU(2)$ class, proving that the resulting operator is non-exact and fortuitous, even when non-Coulombic stable classes are included. This proves the existence of fortuitous BMN cohomology at every $N\geq3$. The construction is an exact algebraic realisation of a nonabelian core with scalar dressing.}
\maketitle

\section{Introduction}\label{sec:intro}

At large $N$, the set of gauge invariant operators is freely generated by the independent single trace operators. At finite $N$, trace identities relate these words. The trace identities have recently been studied systematically in~\cite{deMelloKoch:2025ngs,deMelloKoch:2025cec,deMelloKoch:2025rkw,Caputa:2025ikn,deMelloKoch:2025qeq}; see~\cite{deMelloKoch:2026plq} for a pedagogical review. In the BMN sector of $\mathcal N=4$ super Yang--Mills theory~\cite{Grant:2008sk,Chang:2013fba}, these identities can have a dramatic effect: an operator that is not closed before imposing trace relations may become $Q$-closed at finite rank. Such a cohomology class is called \emph{fortuitous} if it is not the image of any stable large-$N$ cohomology class~\cite{ChangLinOriginal,ChangLin2024,ChoiShape,ChoiMicrostates}. These states are particularly interesting as candidates for black-hole microstates and their existence relies crucially on finite-$N$ dynamics. Explicit fortuitous classes are known at low rank~\cite{ChangLinOriginal,ChoiShape,ChoiMicrostates,ChoiFiniteN,BraneFused,Chang:2024lxt,Chen:2025sum,Kim:2025vup,Belin:2025hsg,Behan:2025hbx,Giusto:2026rpl,Behan:2026her} and the notion has recently been extended to other settings~\cite{ChoiKim:2025rd,Choi:2026faq,Fliss:2026rad}. Related developments extending the understanding of fortuity include~\cite{Budzik:2023vtr,Hughes:2025tdy,Gadde:2025yoa,Chang:2025mqp,Choi:2025bhi,vanLeuven:2025gwr,Chang:2025wgo,Gaikwad:2025ugk,Chryssanthacopoulos:2026mni,KimLeeLeeSong2026,KibeVanLeuvenMathieson,BGRSWZ}. 

The brane-fused construction of~\cite{BraneFused} provides a suggestive precedent. At $SU(3)$ it produces a fortuitous state whose restriction to an $SU(2)\times U(1)$ block contains the $SU(2)$ threshold state accompanied by abelian scalar dressing. This suggests that a nonabelian core responsible for fortuity might survive inside a larger gauge group, while the additional colour directions contribute scalar degrees of freedom. A geometric analysis~\cite{GeometryFortuity} relates the $SU(2)$ threshold obstruction to the vertex of the commuting cone, where the full nonabelian symmetry is restored. This provides a complementary perspective on the core-and-hair picture.

Can the brane-fused core-and-hair mechanism be made explicit for arbitrary $N$? A naive insertion of a two-by-two operator into an $N\times N$ matrix is neither a gauge-invariant construction nor does it provide a proof of cohomological nontriviality. A legitimate construction must produce a multi-trace polynomial operator using the $SU(N)$ letters, and must show that it is closed, non-exact and cannot descend from any stable class, including non-Coulombic gravitons.

We answer the existence question for fortuitous operators at any rank constructively. In the BMN sector of $\mathcal N=4$ super Yang--Mills theory, explicit fortuitous representatives have so far been obtained at small rank, where the cohomology can be computed directly; the present construction supplies explicit representatives for every $N\geq3$. The construction establishes existence. It does not determine the minimal charge of fortuitous states or the completeness of the fortuitous spectrum, and it does not provide a universal origin for all fortuitous states. The guiding idea is the Higgs mechanism. We select an unordered pair of eigenvalues with labels $S=\{i,j\}$, of one of the scalars, say $\phi^1$, to define the block in which the core is supported. If $\phi^1$ has a vacuum expectation value, fields linking the selected ``core'' pair to the remaining spectator colours are massive, with masses given by eigenvalue differences. We define a rank-two projector $P_S$ which projects onto the subspace supporting the core. Using this projector, the most naive attempt to restrict the fields to the core block times dressing fails because the restricted fields do not obey the original $Q$-transformation rules. This can be addressed by correcting the fermions with terms mixing the selected two-dimensional space and its complement. The corrections resemble the elimination of massive off-diagonal multiplets and ensure that the corrected variables obey the $Q$-transformations of the original $SU(2)$ letters. Substituting the corrected letters into the known $SU(2)$ threshold operator produces a $Q$-closed expression associated with the pair $S$. Summing over all pairs defining the core then gives a gauge-invariant polynomial in the original BMN letters.

The argument therefore has two parts. Section~\ref{sec:construction} constructs a gauge-invariant, $Q$-closed polynomial at arbitrary rank using effective two-colour letters built from unrestricted $SU(N)$ fields. Section~\ref{sec:det} then proves that this polynomial represents a nonzero fortuitous class, using a block restriction that preserves stable cohomology and recovers the original $SU(2)$ threshold state. We discuss this second task next.

The second ingredient is a method to detect fortuity. It generalises the $SU(2)\times U(1)$ restriction introduced in~\cite{BraneFused} to show that the $SU(3)$ operator found there is not exact. The idea is to restrict to an $SU(2)$ block together with $N-2$ abelian eigenvalues and extract a chosen power of a parameter controlling their separation. The restriction preserves $Q$-closure and $Q$-exactness and, crucially, \emph{every} stable class maps to a stable class. For our operator the restriction can be calculated exactly: only the pair belonging to the nonabelian block survives. The detector thus returns a nonzero multiple of $O_{24}$. The fortuity of the $SU(2)$ state then rules out both exactness and a stable origin at any higher rank. Thus, the established fortuity of the $SU(2)$ threshold state~\cite{ChangLinOriginal,ChangLin2024,ChoiShape} is an input to our all-rank argument.

Section~\ref{sec:setup} sets up the BMN complex, the $SU(2)$ input and the Higgs-phase picture behind the construction. Section~\ref{sec:result} states the main theorem and its physical meaning. Section~\ref{sec:construction} constructs the operator. Section~\ref{sec:N3} works through the case $N=3$ explicitly; it contains every idea of the general argument with almost no notation. Section~\ref{sec:det} introduces the detector and completes the proof. We conclude in Section~\ref{sec:discussion}. Appendix~\ref{app:proofs} collects the more technical derivations, and Appendix~\ref{app:explicit} gives an explicit multi-trace form of the operator for every $N$, with $N=3$ in complete detail.

\section{Setup and overview}\label{sec:setup}

\subsection{The BMN complex and fortuity}\label{sec:conv}

The letters of the BMN sector are traceless $N\times N$ matrices. The bosonic letters are $\phi^m$ and $f$, while the fermionic letters are $\psi_m$, $m=1,2,3$. We use the normalisation
\begin{equation}\label{eq:Q}
Q\phi^m=0,\qquad Q\psi_m=[\phi^{m+1},\phi^{m+2}],\qquad Qf=\textstyle\sum_m[\phi^m,\psi_m],
\end{equation}
with indices defined mod 3. The standard normalisation includes factors of $i$ that we have absorbed into
the letters. $\cC_N$ denotes the complex of $SU(N)$-invariant polynomials. Since $Q\phi^1=0$, the action of $Q$ extends naturally to rational or locally holomorphic functions of $\phi^1$.

We use the Cartan charges $R_m$ of $SU(3)_R$, under which $\phi^m$ has $R_n=\delta_{mn}$, and $\psi_m$ has $R_m=-\tfrac12$ and $R_{m\pm1}=\tfrac12$, together with $J_1+J_2$, which is $0$ for $\phi^m$, $1$ for $\psi_m$ and $2$ for $f$. We grade operators by the charge $j=2(R_1+R_2+R_3)+3(J_1+J_2)$, and by the fermion number $k= n_\psi+2n_f=J_1+J_2$. It is sometimes convenient to combine these into
\begin{equation}\label{eq:q}
q_m=R_m+\tfrac12(J_1+J_2),
\end{equation}
which counts $1$ for $\phi^m$, $1$ for each $\psi_n$ with $n\neq m$, $0$ for $\psi_m$, and $1$ for $f$. Since $Q$ carries $R_m=\tfrac12$ and $J_1+J_2=-1$, the vector $q=(q_1,q_2,q_3)$ is preserved by $Q$, while $k$ decreases by one.

\paragraph{Gravitons and fortuity.}
Let $\cS$ denote the stable trace algebra, freely generated by single traces of cyclically distinct words in the BMN letters. The action of $Q$ extends to $\cS$ through \eqref{eq:Q}, and evaluation at finite $N$ defines a map $\rho_N:\cS\to\cC_N$ compatible with $Q$. We call the classes in $\im\rho_N^*\subset H(\cC_N)$ \emph{gravitons}, while classes outside this image are \emph{fortuitous}. This definition includes all stable graviton classes, without restricting to the Coulombic sector. Here the Coulombic sector consists of classes detected by evaluating operators on commuting matrices; stable classes not detected in this way are called non-Coulombic.

\subsection{The $SU(2)$ input}\label{sec:input}

The key input to our construction is the $SU(2)$ threshold state, first identified by Chang and Lin~\cite{ChangLinOriginal} and explicitly constructed by Choi et al.~\cite{ChoiShape}. Writing the $SU(2)$ scalar and fermionic letters as $A^m$ and $\Psi_m$, respectively, a convenient trace representation of this operator is
\begin{equation}\label{eq:O24trace}
\begin{aligned}
O_{24}(A;\Psi)=-16i\Big[&\big(\tr\Psi_1A^1-\tr\Psi_2A^2\big)\tr(\Psi_3A^1)\,\tr(\Psi_2\Psi_1^2)\\
&+\big(\tr\Psi_2A^2-\tr\Psi_3A^3\big)\tr(\Psi_1A^2)\,\tr(\Psi_3\Psi_2^2)\\
&+\big(\tr\Psi_3A^3-\tr\Psi_1A^1\big)\tr(\Psi_2A^3)\,\tr(\Psi_1\Psi_3^2)\Big].
\end{aligned}
\end{equation}
The operator contains two scalars and five fermions, with charges $R=(\tfrac32,\tfrac32,\tfrac32)$ and $J_1+J_2=5$, corresponding to $j=24$, $k=5$ and $q=(4,4,4)$. It represents the unique non-graviton class at this threshold~\cite{ChangLinOriginal,ChoiShape}.

The construction of fortuitous states at higher rank relies on two properties of $O_{24}$. First, it is $Q$-closed for arbitrary traceless $2\times2$ matrices $A^m$ and $\Psi_m$ satisfying
\begin{equation}
QA^m=0,\qquad Q\Psi_m=[A^{m+1},A^{m+2}].\label{targetalgebra}
\end{equation}
This closure was established by Choi et al.~\cite{ChoiShape}. Second, the threshold state is a nonzero fortuitous cohomology class, first identified by Chang and Lin~\cite{ChangLinOriginal,ChangLin2024}. We take these established results as inputs. Precisely, the input we use is that $[O_{24}]\neq0$ in $H(\cC_2)$ and $[O_{24}]\notin\im\rho_2^*$, with $\rho_2$ the evaluation map from the \emph{full} stable trace algebra $\cS$. In~\cite{ChangLinOriginal,ChoiShape} the non-graviton property is established by comparison with multi-graviton cohomology; this agrees with $\im\rho_2^*$ because the cohomology of $\cS$ is generated by single-trace gravitons~\cite{ChangLin2024}. In our formulation, fortuity means that the class has no ancestor in the full stable trace cohomology; the higher-rank argument does not restrict stable classes to the Coulombic sector.

The identity $QO_{24}=0$ holds whenever $A^m$ and $\Psi_m$ are replaced by effective letters obeying the same $Q$-transformations (see the substitution argument in Appendix~\ref{app:proofs}). The idea of the construction is to identify effective $2\times2$ letters inside $SU(N)$ that obey \eqref{targetalgebra} and then to use the above $SU(2)$ threshold state to obtain a nontrivial cohomology class at higher rank.

\subsection{Physical picture: Higgsing and the massive multiplets}\label{sec:higgs}

To motivate the construction, consider a diagonal scalar matrix $\phi^1=\mathrm{diag}(\lambda_1,\ldots,\lambda_N)$. For an off-diagonal entry $(i,k)$, the supercharge acts as
\begin{eqnarray}
Q(\psi_3)_{ik}&=&(\lambda_i-\lambda_k)\,(\phi^2)_{ik},\cr\cr
Q(\psi_2)_{ik}&=&-(\lambda_i-\lambda_k)\,(\phi^3)_{ik},\cr\cr
Q f_{ik}&=&(\lambda_i-\lambda_k)\,(\psi_1)_{ik}+\dots .\label{eq:pairs}
\end{eqnarray}
The first two relations are exact for diagonal $\phi^1$; only $Qf_{ik}$ receives further terms, from $[\phi^2,\psi_2]+[\phi^3,\psi_3]$. The eigenvalue difference $\lambda_i-\lambda_k$ plays the role of a mass for the off-diagonal fields, just as in the Higgs phase of gauge theory. When this difference is nonzero, the corresponding linearised $Q$-transformations pair the fields into cohomologically trivial doublets, analogous to BRST doublets~\cite{Barnich2000,Quadri2002}.

Choose two colour directions, $S=\{i,j\}$, and treat the remaining $N-2$ directions as spectators. Our aim is to construct effective $2\times2$ letters on this pair that obey the $Q$-transformations \eqref{targetalgebra} of the $SU(2)$ letters used to define $O_{24}$. The scalar letters and two of the fermions already obey these transformations when restricted to the block. The complication comes from
\begin{equation}
Q\psi_1=[\phi^2,\phi^3].
\end{equation}
The $2\times2$ block of this commutator includes the commutator of the fields within the selected pair, as well as terms involving the spectator colours
\begin{equation}
(Q\psi_1)_{S,S}=[\phi^2_{S,S},\phi^3_{S,S}]+\sum_{c\notin S}\left(\phi^2_{S,c}\phi^3_{c,S}-\phi^3_{S,c}\phi^2_{c,S}\right).
\end{equation}
It is these additional terms that prevent the selected block from behaving like an $SU(2)$ subsystem. Importantly, the unwanted terms can be cancelled by a $Q$-variation. Since the spectator fields are separated from the selected pair by nonzero eigenvalue differences, we can divide by these differences to construct a correction $\mathcal K$, given by
\begin{equation}
\mathcal K_{ij}=\sum_{c\notin S}\frac{(\psi_3)_{ic}(\phi^3)_{cj}+(\psi_2)_{ic}(\phi^2)_{cj}}{\lambda_i-\lambda_c},
\qquad i,j\in S.\label{defK}
\end{equation}
The $Q$-variation of $\mathcal K$ reproduces the spectator terms
\begin{equation}
Q\mathcal K=\sum_{c\notin S}\left(\phi^2_{S,c}\phi^3_{c,S}-\phi^3_{S,c}\phi^2_{c,S}\right).
\end{equation}
Thus, we can define a corrected fermion $\widehat\Psi_1=\psi_{1,S,S}-\mathcal K$, whose $Q$-variation is
\begin{equation}
Q\widehat\Psi_1=[\phi^2_{S,S},\phi^3_{S,S}].
\end{equation}
The spectator contributions cancel exactly, and the resulting transformation reproduces \eqref{targetalgebra} perfectly. The correction $\mathcal K$ involves off-diagonal fermions divided by eigenvalue differences, much like the exchange of massive fields in the Higgs phase. Physically, this correction resembles the exchange of massive off-diagonal multiplets between the selected pair and the spectator colours.

We can now substitute the effective letters into the known $SU(2)$ threshold operator $O_{24}$ to obtain a $Q$-closed operator $O_S$. Since the correction involves inverse eigenvalue differences, $O_S$ is a rational function. Its denominators can be removed by multiplying by a sufficiently high power of
\begin{equation}
\Delta_S=\prod_{a\in S,\;c\notin S}(\lambda_a-\lambda_c).\label{DeltaDef}
\end{equation}
Finally, summing over all choices of the pair $S$ produces an expression symmetric in the eigenvalues, which can be rewritten as a gauge-invariant polynomial in the original matrix letters. Sections~\ref{sec:construction} and~\ref{sec:det} turn this picture into a precise construction. Spectral projectors replace the choice of eigenvectors, while an explicit inverse of the commutator with $\phi^1$ replaces division by eigenvalue differences. The resulting operator is defined without choosing a diagonal basis, i.e. the construction is basis independent.

\section{Main result}\label{sec:result}

In this section we state our main result in the form of a theorem, which is proved in the remainder of the paper. Simply put, we have constructed a family of fortuitous operators in ${\cal N}=4$ super Yang-Mills theory, with gauge group $SU(N)$, for every $N\ge 3$. The statement is in terms of $O_S$, given by evaluating the known fortuitous operator $O_{24}$ on the effective letters and introduced at the end of the last Section, as well as $\Delta_S$ defined in \eqref{DeltaDef}. It also involves a sum over all pairs of eigenvalues on which the core is supported.

\begin{theorem}\label{thm:main}
For every $N\ge3$ and integer $K\ge13$, the construction
\begin{equation}
O_N^{(K)}=\sum_{|S|=2}\Delta_S^K O_S
\end{equation}
defines a gauge-invariant, $Q$-closed polynomial in the BMN letters. Its cohomology class is nonzero and fortuitous: it cannot be obtained from any class in the stable trace algebra, including non-Coulombic gravitons. The operator contains no $f$ letters, has fermion number $k=5$, and carries charges
\begin{equation}
(R_1,R_2,R_3)=\left(\frac32+2K(N-2),\frac32,\frac32\right),\qquad J_1+J_2=5,\qquad j=24+4K(N-2).
\end{equation}
\end{theorem}

\subsection*{Physical implications of the construction}

The theorem implies that the fortuitous $SU(2)$ threshold state can be carried into gauge theories of arbitrarily large rank. The fortuitous operator is an $SU(2)$ nonabelian core dressed by scalar degrees of freedom associated with the remaining $N-2$ colours. The construction initially singles out two colour directions but then sums over all possible pairs of colour directions that can support the core, to produce a gauge-invariant operator with no preferred choice of colours.

The nonabelian core retains its fortuitous character at higher rank. Its nontrivial cohomology depends on the $SU(2)$ trace relations and consequently its $N=2$ core is not the image of any stable multi-graviton class. We prove this, following~\cite{BraneFused}, by restricting the higher-rank operator to a configuration containing an $SU(2)$ block and commuting spectator scalars. This restriction recovers a nonzero multiple of the original $SU(2)$ threshold operator. Since stable classes at rank $N$ restrict to stable classes at rank two, the existence of a stable ancestor for our higher-rank operator would contradict the fortuity of the $SU(2)$ threshold state.

The scalar dressing has a natural interpretation in the Higgs phase. The eigenvalue differences between the selected pair and the spectators determine the masses of the off-diagonal multiplets. The same differences appear in the denominators of the effective letters introduced in Section~\ref{sec:higgs}. Since the eigenvalue differences appear in denominators, we are initially restricted to configurations that have distinct eigenvalues. Multiplying by powers of $\Delta_S$ removes these denominators, extending the construction to a polynomial operator on the full space of matrix configurations. The resulting picture resembles a nonabelian core surrounded by scalar hair.

For fixed $K$, the charge of our operators grows linearly with $N$, well below the expected $O(N^2)$ scaling of macroscopic AdS$_5$ black holes. We therefore do not claim to have constructed macroscopic black-hole microstates. What the theorem establishes is that fortuitous BPS cohomology persists at arbitrarily large rank. Whether more economical dressings or larger nonabelian cores can produce families in the black-hole charge regime is an interesting open question.

The proof proceeds in two stages. Section~\ref{sec:construction} constructs the rational operators $O_S$, proves their $Q$-closure, and shows that multiplication by $\Delta_S^K$, with $K\ge13$, gives a polynomial after summing over pairs. Section~\ref{sec:det} then constructs a restriction map that preserves stable classes and extracts a nonzero multiple of $O_{24}$. Since the latter has no stable ancestor, neither can $O_N^{(K)}$. Section~\ref{sec:N3} makes the general argument concrete through the example of $N=3$.

\section{Construction}\label{sec:construction}

An overview of the construction, in terms of a Higgs-phase picture, was given in Section~\ref{sec:higgs}. In this Section we will develop the details needed to turn this overview into an explicit construction. The idea is to isolate two colour directions inside $SU(N)$ and to construct effective letters that obey the $SU(2)$ supersymmetry transformations. Substituting these into the threshold operator $O_{24}$ gives a $Q$-closed operator at higher rank. The construction has three steps. First, we use spectral projectors to select two colour directions and an inverse commutator to describe the massive off-diagonal fields. The inverse commutator is introduced to determine the correction that cancels unwanted contributions to the commutator algebra \eqref{targetalgebra}, from spectator colours. Second, with the inverse commutator in hand, we correct the fermionic letters to recover the $SU(2)$ transformation rules exactly. Finally, we remove the denominators by multiplying by powers of $\Delta_S$ and sum over all pairs to obtain a gauge-invariant polynomial in the original $SU(N)$ letters.

\subsection{Spectral projectors and propagators}\label{sec:spec}

Let $\lambda_1,\ldots,\lambda_N$ denote the eigenvalues of $\phi^1$. We initially work on the dense open set
\begin{equation}
U=\{\phi^1:\lambda_i\neq\lambda_j\text{ for }i\neq j\},
\end{equation}
where the eigenvalues are distinct. The construction starts by choosing a pair $S=\{1,2\}$. In the end we sum over all possible pairs, so that nothing depends on this original choice and we obtain a gauge invariant operator. All other pairs are treated in exactly the same way. We need to construct two ingredients. The first is a projector that projects onto the two-dimensional subspace defined by our chosen pair. The second is an inverse of the commutator with $\phi^1$, which allows us to correct the fermionic letters by cancelling their unwanted interactions with the spectator colours. In an eigenbasis, this inverse simply divides the off-diagonal fields by the corresponding eigenvalue differences, playing the role of a ``propagator'' for the massive multiplets in the Higgs phase. Both ingredients are constructed directly from $\phi^1$, without choosing specific eigenvectors, i.e. the construction is basis independent. The algebraic derivation of the explicit formulas below is given in Appendix~\ref{app:proofs}.

First consider the construction of the projector. After choosing a pair $S=\{1,2\}$, we define
\begin{equation}
r(x)=\prod_{k\notin S}(x-\lambda_k)=\sum_k r_kx^k .
\end{equation}
The operator $r(\phi^1)$ has a simple action on the eigenvectors of $\phi^1$, which we label by their eigenvalue $|\lambda_i\rangle$. These are right eigenvectors: since $\phi^1$ need not be Hermitian, the projectors constructed below are in general not orthogonal projectors. It is simple to see that
\begin{equation}
r(\phi^1)|\lambda_1\rangle=r(\lambda_1)|\lambda_1\rangle,\qquad
r(\phi^1)|\lambda_2\rangle=r(\lambda_2)|\lambda_2\rangle,\qquad
r(\phi^1)|\lambda_k\rangle=0,\quad k>2 .\label{actr}
\end{equation}
To obtain a projector, we need a second polynomial that compensates for the different factors $r(\lambda_1)$ and $r(\lambda_2)$ on the selected eigenvectors. Towards this end, we define two related functions
\begin{equation}
\tilde v(x)=\frac{(x-\lambda_2)r(\lambda_2)-(x-\lambda_1)r(\lambda_1)}{\lambda_1-\lambda_2},\qquad
{\rm and}\qquad v(x)=\frac{\tilde v(x)}{\Delta_S}\label{vdef}
\end{equation}
where
\begin{equation}
\Delta_S=r(\lambda_1)r(\lambda_2)=\prod_{i\in S,\,k\notin S}(\lambda_i-\lambda_k).
\end{equation}
The factor $\Delta_S$ is nonzero on $U$ and measures the separation between the selected pair and the spectator eigenvalues. Another simple computation proves
\begin{equation}
\tilde{v}(\phi^1)|\lambda_1\rangle=r(\lambda_2)|\lambda_1\rangle,\qquad
\tilde{v}(\phi^1)|\lambda_2\rangle=r(\lambda_1)|\lambda_2\rangle,\qquad
\tilde{v}(\phi^1)|\lambda_k\rangle=0,\quad k>2\,.\label{acttv}
\end{equation}
Thus, combining \eqref{actr} and \eqref{acttv} we have
\begin{equation}
\tilde{v}(\phi^1)r(\phi^1)|\lambda_1\rangle=\Delta_S|\lambda_1\rangle,\qquad
\tilde{v}(\phi^1)r(\phi^1)|\lambda_2\rangle=\Delta_S|\lambda_2\rangle\,.
\end{equation}
We can now construct the spectral projector explicitly. It is immediate to verify that
\begin{equation}
P_S=\frac{\tilde v(\phi^1)r(\phi^1)}{\Delta_S}=v(\phi^1)\,r(\phi^1),\qquad P'_S=1-P_S\label{projectorform}
\end{equation}
are the projectors onto the selected two-dimensional subspace $V_S$ and its complement $V'_S=\ker r(\phi^1)$. This is easily verified by acting on the relevant eigenvectors of $\phi^1$. Although the formula \eqref{vdef} contains a division by the eigenvalue difference $\lambda_1-\lambda_2$, its numerator is antisymmetric under $\lambda_1\leftrightarrow\lambda_2$ and therefore vanishes at $\lambda_1=\lambda_2$. Hence $\lambda_1-\lambda_2$ divides the numerator, and $\tilde v(x)$ is polynomial in $x$ and the eigenvalues, symmetric under $\lambda_1\leftrightarrow\lambda_2$. Thus $P_S$ is a polynomial in $\phi^1$ divided by $\Delta_S$. 

We now construct the inverse commutator. We want to solve the equation
\begin{equation}
[\phi^1,X]=P_SYP'_S,\qquad X=P_SXP'_S
\end{equation}
for $X$. In an eigenbasis of $\phi^1$, the solution is simply
\begin{equation}
X_{ic}=\frac{Y_{ic}}{\lambda_i-\lambda_c},\qquad i\in S,\quad c\notin S.
\end{equation}
Our goal is to obtain this result without choosing an eigenbasis. This can be achieved using the polynomial $r(x)$ introduced above. Since $r(\lambda_c)=0$ for every spectator colour, we have
\begin{equation}
\frac{1}{\lambda_i-\lambda_c}
=\frac{1}{r(\lambda_i)}
\frac{r(\lambda_i)-r(\lambda_c)}{\lambda_i-\lambda_c}.
\end{equation}
The second factor is a polynomial in the eigenvalues. Indeed, writing $r(x)=\sum_k r_k x^k$, we have
\begin{equation}
\frac{a^k-b^k}{a-b}=\sum_{p+q=k-1}a^pb^q\qquad \Rightarrow\qquad\frac{r(a)-r(b)}{a-b}
=\sum_{k=1}^{N-2}r_k\sum_{p+q=k-1}a^pb^q.
\end{equation}
This motivates the matrix expression
\begin{equation}
D_r(Y)=\sum_{k=1}^{N-2}r_k\sum_{p+q=k-1}(\phi^1)^pY(\phi^1)^q.
\end{equation}
Its matrix elements are precisely
\begin{equation}
(D_r(Y))_{ic}=\frac{r(\lambda_i)-r(\lambda_c)}{\lambda_i-\lambda_c}Y_{ic}=\frac{r(\lambda_i)}{\lambda_i-\lambda_c}Y_{ic}.
\end{equation}
Finally, since $v(\lambda_i)=1/r(\lambda_i)$ on the selected pair, the required solution is
\begin{equation}
X=P_Sv(\phi^1)D_r(Y)P'_S.\label{crazynotation}
\end{equation}
Denote the inverse of the commutator $\mathcal M(X)=[\phi^1,X]$ on the selected-to-spectator block by $\mathcal M^{-1}$, so that
\begin{equation}
X=\mathcal M^{-1}(P_SYP'_S).\label{FermCorr}
\end{equation}
The construction is basis independent, and its denominators are controlled by $\Delta_S$.

\subsection{Single-pair operator}\label{sec:OS}

We now use the projectors and inverse commutator to construct an effective $SU(2)$ subsystem inside $SU(N)$. The goal is to obtain scalar and fermionic letters obeying the $Q$-transformations of \eqref{targetalgebra}. Once this is achieved, we can substitute the effective letters into $O_{24}$ to obtain a $Q$-closed operator at rank $N$.

For the selected pair $S$, simplify the notation by writing $P=P_S$ and $P'=P'_S$, and define the scalar matrices acting on the two-dimensional space $V_S$ by
\begin{equation}
B^m=P\phi^mP,\qquad\hat A^m=B^m-\frac12\tr_{V_S}(B^m)\,P.
\end{equation}
The subtraction removes the trace, so that $\hat A^m$ is an $SU(2)$ adjoint matrix. Since $Q\phi^m=0$, we also have $Q\hat A^m=0$. The fermionic letters require more care. Simply restricting the original fermions to the selected block, $P\psi_mP$ works for $\psi_2$ and $\psi_3$, but not for $\psi_1$. Indeed,
\begin{equation}
Q(P\psi_1P)=P[\phi^2,\phi^3]P=[B^2,B^3]+E=[\hat{A}^2,\hat{A}^3]+E,\label{badcomm}
\end{equation}
where
\begin{equation}
E=P\phi^2P'\phi^3P-P\phi^3P'\phi^2P.\label{eq:spectatorE}
\end{equation}
We want the first term on the right hand side of \eqref{badcomm}. The second arises because the matrix product leaves the two-dimensional subspace, passes through a spectator colour and returns. The spectator contributions prevent the original block from obeying the $SU(2)$ rules given in \eqref{targetalgebra}. To correct the fermion, subtract an expression whose $Q$-variation equals $E$. This is achieved using the inverse commutator constructed in Section~\ref{sec:spec}. Define\footnote{Although the matrix products in $\mathcal K$ implicitly sum over all colour indices, by construction the only nonzero matrix elements of $X_m$ are $(X_m)_{i c}$, with $i\in S$ and $c\notin S$, which ensures that only intermediate spectator colours contribute reproducing the expression~\eqref{defK}.}
\begin{equation}
X_m=\mathcal M^{-1}(P\psi_mP'),\qquad\mathcal K=X_3\phi^3P+X_2\phi^2P.\label{eq:correctionK}
\end{equation}
In an eigenbasis of $\phi^1$, the entries of $X_m$ are those of the off-diagonal fermions divided by the eigenvalue differences $\lambda_i-\lambda_c$. Thus $\mathcal K$ contains the same inverse-mass factors that appeared in our Higgs-phase discussion. The key identity is
\begin{equation}
Q\mathcal K=E.
\end{equation}
Consequently, defining
\begin{equation}
\hat\Psi_1=(P\psi_1P-\mathcal K)^0,\qquad\hat\Psi_2=(P\psi_2P)^0,\qquad\hat\Psi_3=(P\psi_3P)^0,
\label{eq:effectivefermions}
\end{equation}
where the superscript $0$ denotes the traceless part on $V_S$, gives the desired $SU(2)$ transformations. The spectator terms cancel exactly.
The corrected letters therefore satisfy
\begin{equation}
Q\hat A^m=0,\qquad Q\hat\Psi_m=[\hat A^{m+1},\hat A^{m+2}],
\end{equation}
with cyclic indices. Indeed, since $P$ commutes with $\phi^1$, the transformations of the second and third fermions immediately give
\begin{equation}
Q(P\psi_2P)=[B^3,B^1]=[\hat{A}^3,\hat{A}^1],\qquad Q(P\psi_3P)=[B^1,B^2]=[\hat{A}^1,\hat{A}^2].
\end{equation}
For the first fermion, we must cancel the spectator contribution $E$ in \eqref{eq:spectatorE}. Since $Q\phi^1=0$, the inverse commutator $\mathcal M^{-1}$ commutes with $Q$. Using the original BMN transformations, we obtain
\begin{equation}
\begin{aligned}
QX_3&=\mathcal M^{-1}\big(P[\phi^1,\phi^2]P'\big)=P\phi^2P',\\
QX_2&=\mathcal M^{-1}\big(P[\phi^3,\phi^1]P'\big)=-P\phi^3P'.
\end{aligned}
\end{equation}
It follows that
\begin{equation}
Q\mathcal K=(QX_3)\phi^3P+(QX_2)\phi^2P=P\phi^2P'\phi^3P-P\phi^3P'\phi^2P=E.
\end{equation}
Therefore
\begin{equation}
Q(P\psi_1P-\mathcal K)=[B^2,B^3]+E-E=[B^2,B^3]=[\hat{A}^2,\hat{A}^3].
\end{equation}
Taking traceless parts does not change these transformations, since commutators are traceless and depend only on the traceless parts of their arguments. This gives effective $2\times2$ letters that satisfy the $SU(2)$ supersymmetry transformations exactly. We can now use them in the threshold operator introduced in Section~\ref{sec:input}.

The original fields $\phi^m,\psi_m$ remain arbitrary traceless $N\times N$ matrices: no block-diagonal restriction has been imposed. Rather, the effective letters are composite expressions supported on $V_S$. In a basis adapted to $V_S\oplus V'_S$, all of them have the form
\begin{equation}
\hat A^m=\begin{pmatrix}A^m_{2\times2}&0\\0&0\end{pmatrix},\qquad
\hat\Psi_m=\begin{pmatrix}\Psi_{m,2\times2}&0\\0&0\end{pmatrix},
\end{equation}
with traceless two-dimensional blocks. Those blocks nevertheless depend on the full $N\times N$ configuration through $P_S$ and, for $\hat\Psi_1$, the spectator correction $\mathcal K$. This construction of effective letters should not be confused with the block restriction used later as a detector in Section~\ref{sec:blockrest}.

For each selected pair $S$, we define
\begin{equation}
O_S=O_{24}(\hat A;\hat\Psi),\label{eq:OSdirect}
\end{equation}
where the effective letters are given above and all traces are taken over $V_S$. To express $O_S$ in terms of the original $N\times N$ letters, define
\begin{equation}
\tau(Y_1\cdots Y_n)=\tr(PY_1P Y_2\cdots P Y_n).
\end{equation}
Writing $v=v(\phi^1)$, introduce
\begin{equation}
\hat{\mathcal K}=v\big[D_r(\psi_3)P'\phi^3+D_r(\psi_2)P'\phi^2\big],\qquad\Psi'_1=\psi_1-\hat{\mathcal K}.
\end{equation}
Since $P\hat{\mathcal K}P=\mathcal K$, the operator $O_S$ is obtained from \eqref{eq:O24trace} by replacing each trace with $\tau$ and $\psi_1$ with $\Psi'_1$, while taking the traceless parts of the effective letters. For example,
\begin{equation}
\tr_{V_S}(\hat\Psi\,\hat A^m)=\tau(\Psi\phi^m)-\frac12\tau(\Psi)\tau(\phi^m),\label{eq:xfactor}
\end{equation}
where $\Psi$ denotes the corresponding corrected fermion. The effective letters obey the $SU(2)$ $Q$-algebra, so the known closure of $O_{24}$ immediately gives
\begin{equation}
QO_S=QO_{24}(\hat A;\hat\Psi)=0.
\end{equation}
The operator is invariant under simultaneous conjugation of the original letters and depends only on the unordered pair $S$. It is initially rational because of the eigenvalue denominators. In the next subsection we remove these denominators and sum over pairs to obtain a gauge-invariant polynomial.

\subsection{Removing denominators and summing over pairs}\label{sec:glob}

We have constructed a $Q$-closed operator $O_S$ for each selected pair $S$. Ideally we would like to write an operator that is a polynomial in the BMN letters. However, $O_S$ contains inverse eigenvalue differences through the projectors and corrected fermions. Our first task is to remove these denominators. In both the projector formula \eqref{projectorform} and the fermionic correction (see the comment after \eqref{FermCorr}) it is manifest that the denominators are powers of\footnote{It is worth pointing out one small subtlety. The definition of $\tilde v(x)$ in \eqref{vdef} appears to contain a denominator $\lambda_1-\lambda_2$. However, its numerator vanishes when $\lambda_1=\lambda_2$, so this denominator cancels. Thus all denominators are accounted for by powers of $\Delta_S$.} $\Delta_S$. Recall that
\begin{equation}
\Delta_S=\prod_{i\in S,\;c\notin S}(\lambda_i-\lambda_c).
\end{equation}
Since $Q\phi^1=0$, we also have $Q\Delta_S=0$. Multiplying $O_S$ by a power of $\Delta_S$ therefore preserves $Q$-closure and removes the denominators. We now determine a sufficient power to remove all denominators.

Each spectral projector $P_S$ contains at most one factor of $\Delta_S^{-1}$. The corrected fermion introduces two further inverse factors: one from the inverse commutator and one from the complementary projector $P'_S$. Every term in $O_{24}$ contains seven letters, at most three of which are $\psi_1$. Projecting the seven letters therefore contributes at most seven inverse factors, while the three possible fermionic corrections contribute at most six more. Thus, we need to cancel at most $7+2\times3=13$ factors of $\Delta_S$. The traceless subtractions do not increase this bound. Consequently,
\begin{equation}
\tilde O_S=\Delta_S^{13}O_S\label{neededpower}
\end{equation}
is polynomial in the original BMN matrix entries and the eigenvalues $\lambda_1,\ldots,\lambda_N$. The power $13$ is sufficient, although we have not proved that it is minimal. 

We have removed the denominators, but $\tilde O_S$ still depends on the eigenvalues used to specify the pair $S$. It is symmetric under exchange of the two eigenvalues in $S$, and under permutations of the spectator eigenvalues. To restore full permutation invariance, we sum over all possible pairs. For any integer $K\geq13$,
define
\begin{equation}
F_K=\sum_{|S|=2}\Delta_S^{K-13}\tilde O_S=\sum_{|S|=2}\Delta_S^K O_S.\label{eq:FK}
\end{equation}
The first expression is polynomial in the original BMN matrix entries and the eigenvalues $\lambda_1,\ldots,\lambda_N$. The second expression is valid on the open set $U$ where the eigenvalues are distinct. The key observation is that $F_K$ is symmetric in the eigenvalues. Indeed, permuting the eigenvalues simply permutes the pairs $S$ appearing in the sum, leaving $F_K$ unchanged. Here the original matrix entries are held fixed; only the eigenvalue labels are permuted.

Why is this useful? Any symmetric polynomial in $\lambda_1,\ldots,\lambda_N$ can be expressed as a polynomial in the elementary symmetric functions $e_k(\lambda)$. But these are precisely the coefficients of the characteristic polynomial of $\phi^1$:
\begin{equation}
\det(x-\phi^1)=\sum_{k=0}^N(-1)^k e_k(\lambda)x^{N-k}.
\end{equation}
Since these coefficients are polynomials in the entries of $\phi^1$, we can replace the symmetric eigenvalue dependence in $F_K$ by expressions involving $\phi^1$ itself. We thus obtain a polynomial $O_N^{(K)}$ in the original $N\times N$ BMN letters, whose expression on $U$ is
\begin{equation}
O_N^{(K)}=\sum_{|S|=2}\Delta_S^K O_S. \label{eq:ONpolynomial}
\end{equation}
Because the projectors and corrected letters transform covariantly under simultaneous conjugation, each term in
this sum is gauge invariant. The resulting polynomial is therefore gauge invariant and can be expressed in terms
of multi-traces~\cite{Procesi}. An explicit multi-trace form of $O_N^{(K)}$, obtained without appeal to the fundamental theorem of symmetric polynomials, is given in Appendix~\ref{app:explicit}, where the case $N=3$ is worked out in full.

Finally, $O_N^{(K)}$ is $Q$-closed. Since $Q\phi^1=0$, we have $Q\Delta_S=0$, and hence on $U$
\begin{equation}
Q(\Delta_S^K O_S)=\Delta_S^K QO_S=0.
\end{equation}
The polynomial $O_N^{(K)}$ is therefore $Q$-closed on $U$. Since $U$ is dense and $O_N^{(K)}$ is polynomial,
this identity extends to all matrix configurations. Each factor $\Delta_S$ has degree $2(N-2)$ in $\phi^1$, while the effective letters carry the charges of their $SU(2)$ counterparts. Consequently, $O_N^{(K)}$ has the charges stated in Theorem~\ref{thm:main}.

We have now constructed a gauge-invariant, $Q$-closed polynomial for every $N\geq3$. It remains to establish
that this operator represents a nonzero fortuitous cohomology class. We first illustrate the key idea at $N=3$, before giving the general proof in Section~\ref{sec:det}.

\section{An explicit example: $N=3$}\label{sec:N3}

The construction is particularly simple for $SU(3)$, where selecting two colours leaves just one spectator. We first write the effective letters explicitly and then show how the block restriction recovers the original $SU(2)$ fortuitous state. The resulting operator is written as an explicit polynomial in the BMN letters in Appendix~\ref{app:explicit}.

\paragraph{Constructing the operator:}
Choose $S=\{1,2\}$, leaving colour $3$ as the spectator. Since $\lambda_1+\lambda_2+\lambda_3=0$, the polynomials of Section~\ref{sec:spec} simplify to
\begin{equation}
r(x)=x-\lambda_3,\qquad\tilde v(x)=-(x+2\lambda_3),\qquad\Delta_S=(\lambda_1-\lambda_3)(\lambda_2-\lambda_3).
\end{equation}
The projector becomes
\begin{equation}
P_S=-\frac{(\phi^1+2\lambda_3)(\phi^1-\lambda_3)}{(\lambda_1-\lambda_3)(\lambda_2-\lambda_3)}.
\end{equation}
In an eigenbasis of $\phi^1$, it is simply $P_S=\operatorname{diag}(1,1,0)$. The fermionic correction is also especially simple. There is only one spectator colour, so the sum in \eqref{defK} contains a single term:
\begin{equation}
\mathcal K_{ij}=\frac{(\psi_3)_{i3}(\phi^3)_{3j}+(\psi_2)_{i3}(\phi^2)_{3j}}{\lambda_i-\lambda_3},\qquad i,j=1,2.
\end{equation}
Using the $Q$-transformations,
\begin{equation}
Q(\psi_3)_{i3}=(\lambda_i-\lambda_3)(\phi^2)_{i3},\qquad Q(\psi_2)_{i3}=-(\lambda_i-\lambda_3)(\phi^3)_{i3},
\end{equation}
we immediately obtain
\begin{equation}
Q\mathcal K_{ij}=(\phi^2)_{i3}(\phi^3)_{3j}-(\phi^3)_{i3}(\phi^2)_{3j}.
\end{equation}
Thus $\mathcal K$ cancels precisely the contribution to $Q\psi_1$ that passes through colour $3$. After taking traceless parts, the effective letters obey the $SU(2)$ $Q$-algebra.

Repeating the construction for the other two pairs and clearing denominators gives
\begin{equation}
O_3^{(K)}=\Delta_{\{1,2\}}^K O_{\{1,2\}}+\Delta_{\{1,3\}}^K O_{\{1,3\}}+\Delta_{\{2,3\}}^K O_{\{2,3\}},
\qquad K\geq13.
\end{equation}
This is the $SU(3)$ example of the gauge-invariant, $Q$-closed polynomial constructed in Section~\ref{sec:construction}. The original $3\times3$ BMN matrices remain unrestricted throughout this construction.

\paragraph{Recovering the $SU(2)$ core.} We now turn to the essential question: does this operator retain the original fortuitous $SU(2)$ state? Following the idea of~\cite{BraneFused}, we evaluate the operator on a configuration containing an $SU(2)$ block and one commuting spectator:
\begin{equation}
\phi^1=(\varphi^1+t\mathbf1_2)\oplus(-2t),\qquad\phi^{2,3}=\varphi^{2,3}\oplus0,\qquad\psi_m=\chi_m\oplus0.
\end{equation}
Here $\varphi^m$ and $\chi_m$ are traceless $2\times2$ BMN letters. We denote this substitution by $\rho_t$; its general-$N$ version is introduced in Section~\ref{sec:blockrest}. The parameter $t$ controls the separation between the nonabelian block and the spectator. The shifts proportional to the identity commute with all other letters, so this restriction preserves the $Q$-transformations. Let the eigenvalues of $\varphi^1$ be $\pm\mu$, where
\begin{equation}
u=\mu^2=\frac12\tr\big((\varphi^1)^2\big).
\end{equation}
The three eigenvalues of $\phi^1$ are then
\begin{equation}
\lambda_+=t+\mu,\qquad\lambda_-=t-\mu,\qquad\lambda_3=-2t.
\end{equation}
We now evaluate the three terms in $O_3^{(K)}$. The crucial observation is that only the pair containing both colours of the $SU(2)$ block survives.

First consider $S=\{+,-\}$. In this case $P_S=\mathbf1_2\oplus0$. There are no off-diagonal fields connecting the block to the spectator, so $\mathcal K=0$. The effective letters reduce to the original $SU(2)$ letters, giving
\begin{equation}
O_{\{+,-\}}=O_{24}(\varphi;\chi).
\end{equation}
The corresponding separation factor is
\begin{equation}
\Delta_{\{+,-\}}=(\lambda_+-\lambda_3)(\lambda_--\lambda_3)=(3t+\mu)(3t-\mu)=9t^2-u.
\end{equation}

What happens to the other two pairs? Consider $S=\{+,3\}$, containing one colour from the $SU(2)$ block and the spectator. Let $\Pi_+$ project onto the eigenvector of $\varphi^1$ with eigenvalue $+\mu$. Because all fermions are
supported in the original $SU(2)$ block, their projections onto this mixed pair are proportional to $\Pi_+$. The correction $\mathcal K$ has the same property. Consequently, after subtracting their traces, all three effective fermions have the form
\begin{equation}
\hat\Psi_m=\hat\zeta_m \left(\Pi_+-\frac12P_S\right),
\end{equation}
where $\hat\zeta_m$ is Grassmann odd; for $m=1$ it includes the contribution of $\mathcal K$. Since $\hat\zeta_m^2=0$, we have $\hat\Psi_m^2=0$. But every term in $O_{24}$ contains a squared fermion. Therefore
\begin{equation}
O_{\{+,3\}}=0.
\end{equation}
The same argument gives $O_{\{-,3\}}=0$. Thus the sum over three pairs collapses to a single contribution:
\begin{equation}
\rho_t\big(O_3^{(K)}\big)=(9t^2-u)^K O_{24}(\varphi;\chi).\label{eq:N3detector}
\end{equation}
Although our derivation has assumed distinct eigenvalues, this is a polynomial identity and therefore extends to coincident eigenvalues as well.

This equation gives a concrete realisation of the core-and-hair picture. The $SU(2)$ fortuitous operator survives unchanged, multiplied by a polynomial describing its separation from the spectator colour. Finally, extracting the highest power of $t$ gives
\begin{equation}
[t^{2K}]\rho_t\big(O_3^{(K)}\big)=9^K O_{24}(\varphi;\chi).
\end{equation}
The block restriction therefore detects a nonzero multiple of the original $SU(2)$ threshold state. This is the key to establishing fortuity. As we show in Section~\ref{sec:det}, the restriction preserves $Q$-exact operators and maps stable cohomology classes to stable classes. Since $O_{24}$ is neither exact nor descended from stable cohomology, the same must be true of $O_3^{(K)}$. The next section extends this argument to arbitrary $N$.

\section{Detector and proof of fortuity}\label{sec:det}

We must now show the operator we have constructed represents a nonzero fortuitous cohomology class. To do this, we generalise the block restriction illustrated at $N=3$ in the previous section. We restrict the $SU(N)$ matrices to a configuration containing an $SU(2)$ block and $N-2$ commuting spectator colours. We then show that this restriction extracts the original threshold operator $O_{24}$, multiplied by a nonzero polynomial in the scalar eigenvalues. The restriction preserves $Q$-exact operators and maps stable classes at rank $N$ to stable classes at rank two. Since $O_{24}$ is neither exact nor descended from a stable class, the same must hold for our higher-rank operator.

\subsection{Block restriction}\label{sec:blockrest}

To detect the $SU(2)$ core of our higher-rank operator, restrict the $SU(N)$ matrices to a $2\times2$ block accompanied by $N-2$ commuting spectator colours. The restriction recovers the original threshold operator $O_{24}$. More importantly, it preserves both $Q$-exactness and the property of having a stable ancestor, making it possible to establish fortuity at higher rank. Choose constants $b_3,\ldots,b_N$ satisfying
\begin{equation}
2+\sum_{k=3}^N b_k=0,\label{eq:bsum}
\end{equation}
so that the scalar matrices remain traceless. We introduce an auxiliary parameter $t$, which controls the separation between the $SU(2)$ block and the spectators, and define
\begin{equation}
\begin{aligned}
\phi^1&\longmapsto(\varphi^1+t\mathbf1_2)\oplus t\,\mathrm{diag}(b_3,\ldots,b_N),\qquad
\phi^{2,3}\longmapsto\varphi^{2,3}\oplus0,\\
\psi_m&\longmapsto\chi_m\oplus0,\qquad f\longmapsto g\oplus0,
\end{aligned}
\end{equation}
where $(\varphi,\chi,g)$ are the original $SU(2)$ BMN letters. Denote this substitution by
\begin{equation}
\rho_t:\mathcal C_N\longrightarrow\mathcal C_2[t].
\end{equation}
The restriction is compatible with supersymmetry. All the matrices are block diagonal, so their commutators are supported entirely in the $2\times2$ block. The terms proportional to the identity and the diagonal spectator entries commute with everything else and therefore drop out of the $Q$-transformations. Consequently,
\begin{equation}
\rho_t(QO)=Q\rho_t(O)
\end{equation}
for any $O$. So, a $Q$-closed operator remains closed after restriction, while a $Q$-exact operator remains exact. Since $Qt=0$, we can also extract any coefficient in the expansion
\begin{equation}
\rho_t(O)=\sum_{h\ge0}t^h\beta_h(O).
\end{equation}
Each coefficient separately respects the action of $Q$
\begin{equation}
\beta_h(QO)=Q\beta_h(O).
\end{equation}
Thus the maps $\beta_h:\mathcal C_N\to\mathcal C_2$ preserve $Q$-closure and $Q$-exactness. The crucial additional property is that these maps also preserve stable cohomology classes. To see why, recall that a stable operator is built from products of traces, without imposing finite-$N$ trace relations. Under the block restriction, each trace separates into a contribution from the $2\times2$ block and a contribution from the commuting spectators:
\begin{equation}\label{eq:blocktrace}
\tr_N\prod_i\left((L_i^{(2)}+a_i\mathbf1_2)\oplus\mathrm{diag}(b_{i,k})\right)=
\tr_2\prod_i(L_i^{(2)}+a_i\mathbf1_2)+\sum_k\prod_i b_{i,k}.
\end{equation}
To see why this also holds \emph{before} finite-rank evaluation, write $T(W)$ for the formal single-trace generator associated with a cyclic word $W$ in the stable algebra $\cS$, and let $\widehat\rho_t:\cS\to\cS[t]$ denote the block substitution applied to formal traces. For example,
\begin{equation}\label{eq:stable-example}
\widehat\rho_t\big(T(\phi^1\phi^2\psi_2)\big)
=T(\varphi^1\varphi^2\chi_2)+t\,T(\varphi^2\chi_2).
\end{equation}
The $T$'s on the right are \emph{formal} trace symbols, not traces of $2\times2$ matrices. This rule extends linearly to cyclic words and multiplicatively to products of formal traces. Words built only from $\phi^1$ also acquire a spectator term,
\begin{equation}\label{eq:stable-phi1}
\widehat\rho_t\big(T((\phi^1)^n)\big)=T\big((\varphi^1+t)^n\big)+t^n\sum_{k=3}^N b_k^n,
\end{equation}
where the first term is expanded binomially and the empty word that appears is assigned the value $T(\emptyset)=2$. It involves no finite-$N$ trace identity. The constraint \eqref{eq:bsum} ensures that the traceless single-letter relation is respected, including for $T(\phi^1)$.

The map $\widehat\rho_t:\cS\to\cS[t]$ thus obtained satisfies
\begin{equation}\label{eq:stable-square}
\rho_t\circ\rho_N=\rho_2\circ\widehat\rho_t,
\qquad Q\widehat\rho_t=\widehat\rho_t Q.
\end{equation}
The first identity follows by multiplying block matrices; the second follows because the shifts are central and $Qt=0$, so the BMN commutators transform correctly, and because the $Q$-variation of any word containing a letter other than $\phi^1$ again contains such a letter, so it receives no spectator contribution, while words built only from $\phi^1$ are $Q$-closed. Both identities hold at the level of formal trace words, prior to imposing any finite-rank identities. Consequently, if $G\in\cS$ obeys $QG=0$, then every coefficient $[t^h]\widehat\rho_t(G)$ is itself a $Q$-closed element of $\cS$, and
$\beta_h(\rho_N(G))=\rho_2([t^h]\widehat\rho_t(G))$. This is the additional property needed beyond preservation of $Q$-exactness. In terms of the evaluation maps introduced in Section~\ref{sec:setup}, this means
\begin{equation}
\beta_h^*\big(\mathrm{im}\,\rho_N^*\big)\subseteq\mathrm{im}\,\rho_2^*.
\end{equation}
Further details of the map $\widehat\rho_t$ are collected in Appendix~\ref{app:proofs}.

This conclusion does not use a classification of stable cohomology. In particular, it includes non-Coulombic graviton classes, which need not be detected by evaluating operators on commuting matrices. The restriction preserves both exact and stable classes. If its action on our higher-rank operator produces a nonzero multiple of the fortuitous $SU(2)$ threshold state, the higher-rank operator cannot be exact or have a stable ancestor.

\subsection{Exact evaluation of the detector}\label{sec:eval}

We can evaluate the restriction of our operator exactly. Only the pair containing the entire $SU(2)$ block contributes, while all other pairs vanish. To see this, write the eigenvalues of $\varphi^1$ as $\pm\mu$, where
\begin{equation}
u=\frac12\tr(\varphi^1)^2=\mu^2.
\end{equation}
The eigenvalues of the full $SU(N)$ matrix $\phi^1$ are then
\begin{equation}
\lambda_\pm=t\pm\mu,\qquad \lambda_k=tb_k,\quad k=3,\ldots,N.
\end{equation}
There are three possibilities for the sum over pairs. The reason the mixed pairs vanish is especially simple: after restriction, their effective fermions all lie along the same traceless matrix in the chosen two-dimensional subspace. Each is a Grassmann-odd coefficient times that matrix, so its square vanishes. Every term of $O_{24}$ contains such a squared fermion. The calculation below makes this observation precise.

\medskip
\noindent
\textit{The pair contains the full $SU(2)$ block:} For $S=\{+,-\}$, the projector is simply $P=\mathbf1_2\oplus0$. Since none of the fermions connect this block to the spectators, $P\psi_mP'=0$, and the fermionic correction vanishes. The effective letters are the original $SU(2)$ letters, giving
\begin{equation}
O_{\{+,-\}}=O_{24}(\varphi;\chi).
\end{equation}
The corresponding eigenvalue factor is
\begin{equation}
\Delta_{\{+,-\}}=\prod_{k=3}^N(\lambda_+-\lambda_k)(\lambda_--\lambda_k)=\prod_{k=3}^N\big(t^2(1-b_k)^2-u\big).
\end{equation}

\medskip
\noindent
\textit{Pairs containing one $SU(2)$ direction and one spectator:} Consider $S=\{\pm,k\}$. Let $\Pi_\pm$ project onto one eigenvector of $\varphi^1$, and let $\Pi_k$ project onto the spectator eigenvector. Then $P=\Pi_\pm+\Pi_k$. All fermions are supported in the original $SU(2)$ block, so their projections onto the selected pair are proportional to the same rank-one projector:
\begin{equation}
P\psi_m P=\zeta_m\Pi_\pm.
\end{equation}
This remains true for the corrected fermion. Indeed, $P'=\Pi_\mp+\sum_{l\neq k}\Pi_l$, and since the fermions are supported in the $SU(2)$ block, $P\psi_mP'=\Pi_\pm\psi_m\Pi_\mp$. The inverse commutator divides by $\lambda_\pm-\lambda_\mp=\pm2\mu$, so $X_m=\pm\Pi_\pm\psi_m\Pi_\mp/(2\mu)$. Because $\phi^{2,3}$ have no spectator entries, $\Pi_\mp\phi^{2,3}\Pi_k=0$, and therefore
\begin{equation}
\mathcal K=\pm\frac{1}{2\mu}\,\Pi_\pm\big(\psi_3\Pi_\mp\phi^3+\psi_2\Pi_\mp\phi^2\big)\Pi_\pm\ \propto\ \Pi_\pm.
\end{equation}
Hence $P\psi_1P-\mathcal K$ is again a Grassmann-odd multiple of $\Pi_\pm$. Write the projected corrected fermions as $\hat\zeta_m\Pi_\pm$, where $\hat\zeta_m=\zeta_m$ for $m=2,3$ and $\hat\zeta_1$ includes the contribution of $\mathcal K$. Each $\hat\zeta_m$ is Grassmann odd, and hence $\hat\zeta_m^2=0$. Denoting the corrected fermions $\Psi'_1,\psi_2,\psi_3$ collectively by $\Psi_m$,
\begin{equation}
\tau(\Psi_b\Psi_a^2)=\hat\zeta_b\hat\zeta_a^2\tr\Pi_\pm=0.
\end{equation}
Every term in the threshold operator \eqref{eq:O24trace} contains such a factor. Therefore $O_{\{\pm,k\}}=0$.

\medskip
\noindent
\textit{Pairs containing only spectators:} For $S\subset\{3,\ldots,N\}$, the fermionic letters vanish identically after projection, so these pairs also give $O_S=0$.

Thus only the pair containing the full $SU(2)$ block survives and we have the exact identity
\begin{equation}\label{eq:block}
\rho_t\big(O_N^{(K)}\big)=\prod_{k=3}^N\big(t^2(1-b_k)^2-u\big)^K\,O_{24}(\varphi;\chi).
\end{equation}
Although the calculation used distinct eigenvalues, both sides are polynomials in the matrix letters, $t$ and the spectator parameters $b_k$. The identity therefore extends to all configurations by polynomial continuity. Thus the higher-rank operator retains the original $SU(2)$ threshold state, multiplied by a polynomial describing its separation from the spectator colours. We can isolate this core by extracting the highest power of $t$. Since there are $N-2$ spectators, that power is $h=2K(N-2)$, and we find
\begin{equation}
\beta_h\big(O_N^{(K)}\big)=\left[\prod_{k=3}^N(1-b_k)^{2K}\right]O_{24}(\varphi;\chi).
\end{equation}
The prefactor is nonzero whenever $b_k\neq1$. For example, choosing all spectator parameters equal,
\begin{equation}
b_k=-\frac{2}{N-2},
\end{equation}
gives
\begin{equation}
\beta_h\big(O_N^{(K)}\big)=\left(\frac{N}{N-2}\right)^{2K(N-2)}O_{24}(\varphi;\chi).
\end{equation}
This choice is allowed because the polynomial identity holds even when spectator eigenvalues coincide.

We have therefore constructed a restriction that extracts a nonzero multiple of the $SU(2)$ threshold operator from our higher-rank state. The same restriction preserves $Q$-exact operators and stable cohomology classes. Since $O_{24}$ is neither exact nor descended from a stable class, the identity above is the essential ingredient in proving fortuity at every $N$.

\subsection{Establishing fortuity}

We can now complete the proof of our main result. Section~\ref{sec:glob} established that $O_N^{(K)}$ is a gauge-invariant, $Q$-closed polynomial with the stated charges. It remains to show that its cohomology class is fortuitous.

Suppose, to the contrary, that $[O_N^{(K)}]$ descended from a stable cohomology class, including the possibility that it is $Q$-exact. Since the detector preserves stable classes, the identity above would imply that a nonzero multiple of $[O_{24}]$ also has a stable ancestor. But this contradicts the fortuity of the $SU(2)$ threshold state, reviewed in Section~\ref{sec:input}~\cite{ChangLinOriginal,ChangLin2024,ChoiShape}. Thus $[O_N^{(K)}]$ is nonzero and fortuitous for every $N\ge3$ and $K\ge13$, completing the proof.

\section{Discussion}\label{sec:discussion}

The clearest precursor of our result is the brane-fused $SU(3)$ operator of~\cite{BraneFused}. Its $SU(2)\times U(1)$ restriction reveals the threshold state $O_{24}$ together with scalar dressing. This raises an obvious question: can the same fortuitous $SU(2)$ core be embedded in an arbitrary number of colours, without losing its cohomological distinction from stable multi-graviton states? Our construction answers this question. Rather than guess higher-rank operators from particular fusion products, we use spectral projectors to select each possible two-colour subspace and correct the fermions so that the effective letters obey the $SU(2)$ supersymmetry transformations exactly. Removing the eigenvalue denominators and summing over pairs produces an invariant polynomial. The detector then returns $O_{24}$ multiplied by an explicit scalar separation factor. This extends the core-and-hair mechanism seen at $SU(3)$ to every $N\geq3$, although it does not identify our operators with the particular brane-fused representatives.

The geometry of fortuity offers a complementary way of understanding the surviving core~\cite{GeometryFortuity}. At $SU(2)$, the lifting obstruction of $O_{24}$ is supported at the vertex of the commuting cone, where the full nonabelian gauge symmetry is restored. At $SU(3)$, the geometric analysis relates the dressed core to the $2+1$ joint-eigenvalue collision stratum, while distinguishing this from a proof of support for the full obstruction. Our block detector similarly singles out an $SU(2)$ core with commuting scalar spectators. These observations suggest that the enhanced-symmetry strata of the commuting Hironaka cover may help organise analogous cores at higher rank. This is an interpretation, not an ingredient of our proof: the detector does not establish the support of the all-$N$ class, and the polynomial representative is defined also for separated eigenvalues.

The ``nonabelian core'' is the part of the construction that rules out a stable ancestor. The proof of fortuity comes from $O_{24}$, which is closed at finite rank and is not obtained from stable trace cohomology. The higher-rank detector preserves this distinction. It is tempting to call the spectator eigenvalues ``hair''. That language has a limited meaning: they supply commuting scalar directions and an explicit polynomial separation factor. We did not prove that the resulting states are tensor products of an independent core Hilbert space and a hair Hilbert space. Further, summing over pairs is essential to restoring the full gauge symmetry.

The stable sector entering the proof is the \emph{complete} stable trace cohomology. It is not restricted to diagonal, Coulombic or otherwise special multi-graviton representatives. Thus possible non-Coulombic graviton sectors cannot invalidate the non-exactness test: any stable ancestor, of whatever form, would restrict to a stable $SU(2)$ ancestor and contradict the known fortuity of $O_{24}$~\cite{ChangLinOriginal,ChangLin2024,ChoiShape}. This avoids a potential loophole in arguments that classify multi-graviton operators using only the commuting locus.

Finite-$N$ cohomology is of interest partly because the stable multi-graviton description is not expected to capture the entire BPS spectrum at black-hole charges. Fortuitous classes supply a sharply defined way in which the finite-rank Hilbert space can contain protected states that are invisible to stable cohomology. Our theorem shows that this mechanism persists at every rank, rather than appearing only as an accident of the small matrices where explicit calculations are feasible.

For fixed $K$ the constructed charge is $j=24+4K(N-2)$, so our family lives at $O(N)$ charge and may be more naturally interpreted as a \emph{fortuitous giant graviton brane state}, rather than a state with $O(N^2)$ charge relevant to the entropy of macroscopic black holes. Further, the constructed cohomology class does not imply an exponentially large number of states. So, the relevance for black holes is structural: the all-rank family demonstrates a stable mechanism for producing genuinely finite-$N$ protected sectors. A natural next question is whether many inequivalent cores, their allowed dressings, or more economical removal of poles can generate sufficiently rich cohomology at higher charges. Understanding the support of the resulting lifting obstructions and whether it follows the enhanced-symmetry strata suggested above, would be a valuable separate test. These questions connect the constructive theorem to the study of the spectrum.

The hair is not optimised. At $N=3$ the theorem gives $j=24+52=76$ for $K=13$. The known $SU(3)$ class $O^3_{30}$ of~\cite{BraneFused} has $j=30$, i.e. three units of abelian hair. Its $SU(2)\times U(1)$ restriction contains $-\tfrac{648}{5}\,\hat\phi^{(i}\hat\phi^j \hat\phi^{k)}\,O_{24}$, where $\hat\phi^i$ denotes the abelian $U(1)$ component of $\phi^i$ in this restriction~\cite{BraneFused}, which has the same core-times-dressing structure that our detector reads off. Note, however, that our family has $j=24+4K$ at $N=3$, with hair carried only by even powers of $\phi^1$, so no member of the family coincides with $O^3_{30}$, whose hair is cubic and carries general $SU(3)_R$ indices. The exponent $K\ge13$ is a sufficient bound from counting denominators, not a minimal dressing. We do not claim that it is sharp. Lowering the hair, ideally down to the value realised by $O^3_{30}$, may require $Q$-exact corrections that soften the poles of $O_S$; we have not investigated this. We will return to this elsewhere. The construction also does not classify all fortuitous families or determine whether two different choices of core yield linearly independent classes. For another $SU(2)$ core with $n$ letters and maximal $\psi_1$-degree $n_{\psi_1}$, the same denominator count suggests $K\ge n+2n_{\psi_1}$, but the mixed-pair vanishing must be established separately; for $O_{24}$ it follows from the explicit trace structure used above.

The all-rank proof is analytic. It uses the known $SU(2)$ threshold operator, elementary matrix identities, polynomial continuation and an exact core detector. The brane-fused work motivates the core-and-hair construction and supplies its first higher-rank example. Our construction shows that a protected, intrinsically finite-rank nonabelian core can be promoted to an invariant fortuitous operator for every larger gauge group.

\acknowledgments
The work was supported by a start-up research fund of Huzhou Normal University (RdMK, MK), a Zhejiang Province talent award (RdMK), a Changjiang Scholar award (RdMK), the Korea NRF Grant No.RS-2025-25414114 (MK), and the Brain Pool Program funded by the Ministry of Science and ICT through the National Research Foundation of Korea (RS-2025-25457100) (SL).

\appendix
\section{Technical derivations}\label{app:proofs}

This appendix collects algebraic details deferred from the main text. We first explain why substituting the effective letters into the $SU(2)$ threshold operator preserves $Q$-closure. The operator $O_{24}(A;\Psi)$ is $Q$-closed for arbitrary traceless $2\times2$ matrices satisfying
\begin{equation}
QA^m=0,\qquad Q\Psi_m=[A^{m+1},A^{m+2}],
\end{equation}
with cyclic indices. Since $Q$ acts on products through the graded Leibniz rule, its action on any polynomial in these letters is determined by their individual transformations. Replacing $A^m$ and $\Psi_m$ by effective traceless letters obeying the same rules therefore preserves the identity $QO_{24}=0$. In particular, the effective letters constructed in Section~\ref{sec:OS} satisfy these transformations exactly, so
\begin{equation}
QO_S=QO_{24}(\hat A;\hat\Psi)=0.
\end{equation}
More precisely, the closure identity $QO_{24}=0$ holds in the free supercommutative algebra generated by the matrix entries of $A^m$ and $\Psi_m$, with $Q$ acting as an odd derivation. It therefore holds for $2\times2$ matrices with entries in any supercommutative algebra on which $Q$ acts by these rules. The effective letters become $2\times2$ matrices after a local choice of basis of $V_S$; since $O_{24}$ is built from traces, the result does not depend on this choice. This is the substitution principle used in the construction. It applies because the closure identity for $O_{24}$ holds for arbitrary $SU(2)$ letters, not just for a particular matrix configuration.

\paragraph{A useful trace identity .}

In \eqref{eq:O24trace} we may replace the fermions by non-traceless odd matrices, provided each
factor $\tr(\Psi A)$ is read as $\tr(\Psi A^0)$ with $A^0$ traceless. Indeed, if
$\Psi=\Psi^0+\tfrac12 s\,\mathbf 1$ with $s$ odd, then $\Psi^2=(\Psi^0)^2$, because
$s\Psi^0+\Psi^0 s=0$ and $s^2=0$. Also $\tr\big((\Psi^0)^2\big)=0$, because $(\Psi^0)^2$ is traceless.
Hence $\tr(\Psi_b\Psi_a^2)=\tr(\Psi_b^0(\Psi_a^0)^2)$.

\paragraph{Spectral projector.}
In an eigenbasis of $\phi^1$, the projector $P_S=v(\phi^1)r(\phi^1)$ acts by multiplication by $v(\lambda_i)r(\lambda_i)$ on the $i$th eigenvector. By construction, this equals $1$ for $i=1,2$, while $r(\lambda_i)=0$ for all spectator eigenvalues. Thus $P_S=\mathrm{diag}(1,1,0,\ldots,0)$, $P'_S=1-P_S$, which immediately establishes the required projector properties.

\paragraph{Inverse commutator.}
In an eigenbasis of $\phi^1$, $[\phi^1,X]=P_SYP'_S$ reduces to
\begin{equation}
(\lambda_i-\lambda_{c})X_{i c}=Y_{i c},\qquad i\in S,\quad c\notin S.
\end{equation}
Since the eigenvalues are distinct on $U$, this has the unique solution
\begin{equation}
X_{i c}=\frac{Y_{i c}}{\lambda_i-\lambda_{c}}.
\end{equation}
On the other hand, the definition of $D_r$ gives
\begin{equation}
(D_r(Y))_{ic}=\frac{r(\lambda_i)-r(\lambda_{c})}{\lambda_i-\lambda_{c}}Y_{ic}=\frac{r(\lambda_i)}{\lambda_i-\lambda_{c}}Y_{ic},
\end{equation}
where we used $r(\lambda_{c})=0$ for spectator eigenvalues. Since $v(\lambda_i)r(\lambda_i)=1$ on the selected pair, we recover
\begin{equation}
X=P_Sv(\phi^1)D_r(Y)P'_S.
\end{equation}

\paragraph{Restriction of stable classes.}
We now verify that the block restriction maps stable classes at rank $N$ to stable classes at rank two. Consider any single trace $T(w)$ in the stable algebra. Under the block substitution of Section~\ref{sec:blockrest}, the trace separates into a contribution from the $2\times2$ block and a sum over the spectator eigenvalues, as in \eqref{eq:blocktrace}. Expanding the first contribution in powers of $t$ gives a polynomial in stable $SU(2)$ traces, while the spectator contribution gives an ordinary polynomial in $t$.

This defines the restriction $\widehat\rho_t:\cS\to\cS[t]$ of Section~\ref{sec:blockrest}, which extends to products of traces. Here the trace of the empty word is assigned the value $\tr_2\mathbf 1=2$, as in \eqref{eq:stable-phi1}, and condition \eqref{eq:bsum} ensures that $\widehat\rho_t(\tr\phi^1)$ vanishes, consistent with tracelessness. Formal compatibility with $Q$ holds because the $Q$-variation of any word containing a letter other than $\phi^1$ again contains such a letter, so it receives no spectator contribution, while words built only from $\phi^1$ are $Q$-closed. Together with the centrality of the scalar shifts and $Qt=0$, this gives $Q\widehat\rho_t=\widehat\rho_t Q$ in \eqref{eq:stable-square}, and hence $\beta_h(\rho_NG)=\rho_2\big([t^h]\widehat\rho_t(G)\big)$ for every $Q$-closed $G\in\cS$. Thus the detector maps every stable cohomology class at rank $N$ to a stable class at rank two, without requiring any classification of graviton operators.

\paragraph{Denominator bound.} The count showing that $\Delta_S^{13}O_S$ is polynomial is given in Section~\ref{sec:glob}. Appendix~\ref{app:explicit} makes it explicit, by writing $\Delta_S^{13}O_S$ as a polynomial in two variables in which every monomial carries exactly thirteen powers of $\Delta_S$.

\section{An explicit multi-trace representative}\label{app:explicit}

Section~\ref{sec:glob} established that $O_N^{(K)}$ is a polynomial in the BMN letters by appealing to the fundamental theorem of symmetric polynomials. That argument proves existence but does not exhibit the polynomial. In this appendix we give a direct construction, valid for every $N$, which produces $O_N^{(K)}$ as an explicit expression in traces. We work out $N=3$ completely. By ``completely'' we mean that every ingredient is given in closed form, so that the operator can be evaluated on any configuration. We do not expand the final traces into individual words, which is a long list of terms. An accompanying Mathematica notebook, posted as an ancillary file, verifies the matrix identities used below and implements the $N=3$ formula.

\subsection{Pair data as polynomials in two variables}

Write the characteristic polynomial of $\phi^1$ as
\begin{equation}
{\mathcal X}(t)=\det(t-\phi^1)=\sum_{k=0}^N(-1)^ke_k\,t^{N-k},\qquad e_1=0,
\end{equation}
whose coefficients are polynomials in the entries of $\phi^1$, and hence multi-traces of $\phi^1$. In place of the two selected eigenvalues introduce two variables $x$ and $y$, and define $r_{xy}(t)$ and $\varrho_{xy}(t)$ by polynomial division,
\begin{equation}
{\mathcal X}(t)=(t-x)(t-y)\,r_{xy}(t)+\varrho_{xy}(t),\qquad \deg\varrho_{xy}\leq1.\label{eq:rxy}
\end{equation}
The coefficients of $r_{xy}(t)=\sum_k r_k(x,y)\,t^k$ are polynomials in $x$, $y$ and the $e_k$, symmetric under $x\leftrightarrow y$. When $x=\lambda_a$ and $y=\lambda_b$ are two distinct eigenvalues of $\phi^1$, the remainder vanishes and $r_{xy}$ reduces to the polynomial $r$ of Section~\ref{sec:spec} for the pair $S=\{a,b\}$. With this definition set
\begin{equation}
\Delta(x,y)=r_{xy}(x)\,r_{xy}(y),\qquad
\tilde v_{xy}(t)=\frac{(t-y)\,r_{xy}(y)-(t-x)\,r_{xy}(x)}{x-y}.
\end{equation}
The numerator of $\tilde v_{xy}$ is antisymmetric under $x\leftrightarrow y$, so the division is exact and $\tilde v_{xy}(t)$ is a polynomial in $t$, $x$ and $y$, symmetric in $x$ and $y$. Finally define the matrix polynomials
\begin{equation}
\mathcal P=\tilde v_{xy}(\phi^1)\,r_{xy}(\phi^1),\qquad
\mathcal P'=\Delta(x,y)-\mathcal P,\qquad
D_{xy}(Y)=\sum_{k=1}^{N-2}r_k(x,y)\sum_{p+q=k-1}(\phi^1)^pY(\phi^1)^q .
\end{equation}
Comparing with Section~\ref{sec:spec}, at $(x,y)=(\lambda_a,\lambda_b)$ we have
\begin{equation}
\Delta=\Delta_S,\qquad \mathcal P=\Delta_SP_S,\qquad \mathcal P'=\Delta_SP'_S,\qquad \tilde v_{xy}(\phi^1)=\Delta_S\,v(\phi^1),\qquad D_{xy}=D_r .
\end{equation}
Every quantity on the left is a polynomial in $x$, $y$ and the matrix entries, with no denominators.

\subsection{The single-pair operator as a polynomial}

Define the scaled letters
\begin{equation}
\begin{aligned}
L^m&=\mathcal P\,\phi^m,\qquad F_2=\mathcal P\,\psi_2,\qquad F_3=\mathcal P\,\psi_3,\\
F_1&=\mathcal P\Big[\Delta^2\psi_1-\tilde v_{xy}(\phi^1)\big(D_{xy}(\psi_3)\,\mathcal P'\phi^3+D_{xy}(\psi_2)\,\mathcal P'\phi^2\big)\Big].
\end{aligned}\label{eq:scaledletters}
\end{equation}
At $(x,y)=(\lambda_a,\lambda_b)$ these become $L^m=\Delta_S P\phi^m$, $F_{2,3}=\Delta_S P\psi_{2,3}$ and $F_1=\Delta_S^3P\Psi'_1$, with $\Psi'_1=\psi_1-\hat{\mathcal K}$ as in Section~\ref{sec:OS}. A trace of a product of scaled letters is therefore a $\tau$-trace multiplied by a known power of $\Delta_S$: one power for each $L^m$, $F_2$ or $F_3$, and three for each $F_1$.

Now take the trace form \eqref{eq:O24trace} of $O_{24}$ and make the replacements
\begin{equation}
\tr(\Psi_kA^m)\ \to\ \tr(F_kL^m)-\tfrac12\tr(F_k)\tr(L^m),\qquad
\tr(\Psi_b\Psi_a^2)\ \to\ \tr(F_bF_aF_a),
\end{equation}
multiplying each monomial by $\Delta^{6-2n_1}$, where $n_1\leq3$ is the number of $\psi_1$'s in that monomial. Call the result $G_{13}(x,y)$. A monomial with $n_1$ factors of $\psi_1$ carries $2+(5-n_1)+3n_1=7+2n_1$ powers of $\Delta_S$ from its letters, so after the weight every monomial carries exactly $\Delta_S^{13}$. By \eqref{eq:xfactor} and the trace identity of Appendix~\ref{app:proofs}, which allows non-traceless fermions in $\tr(\Psi_b\Psi_a^2)$, this gives
\begin{equation}
G_{13}(\lambda_a,\lambda_b)=\Delta_S^{13}\,O_S .
\end{equation}
By construction $G_{13}(x,y)$ is a polynomial in $x$, $y$ and the BMN letters, symmetric under $x\leftrightarrow y$, whose $x$- and $y$-independent ingredients are traces of products of the letters and of powers of $\phi^1$. This gives an explicit form of the bound $K\geq13$ of Section~\ref{sec:glob}. For $K\geq13$ set $G_K=\Delta^{K-13}G_{13}$.

\subsection{Summing over pairs}

Expand $G_K(x,y)=\sum_{i,j}g_{ij}\,x^iy^j$, where $g_{ij}=g_{ji}$ are multi-trace polynomials in the BMN letters. On $U$,
\begin{equation}
O_N^{(K)}=\sum_{a<b}G_K(\lambda_a,\lambda_b)=\frac12\sum_{a\neq b}G_K(\lambda_a,\lambda_b).
\end{equation}
Using $\sum_a\lambda_a^i=\tr\left((\phi^1)^i\right)$, the sum over ordered pairs and the subtraction of the terms with $a=b$ give
\begin{equation}
O_N^{(K)}=\frac12\sum_{i,j}g_{ij}\Big[\tr\big((\phi^1)^i\big)\,\tr\big((\phi^1)^j\big)-\tr\big((\phi^1)^{i+j}\big)\Big].\label{eq:doubletrace}
\end{equation}
Both sides are polynomials, so the identity, derived on $U$, holds everywhere. Since every $\lambda_a$ is a root of ${\mathcal X}$, one may first reduce $G_K(x,y)$ modulo ${\mathcal X}(x)$ and ${\mathcal X}(y)$; then only $i,j\leq N-1$ appear and \eqref{eq:doubletrace} is a finite sum. Equation \eqref{eq:doubletrace} is a manifestly gauge-invariant multi-trace polynomial that makes no reference to eigenvalues. It gives an explicit representative of the class of Theorem~\ref{thm:main} for every $N$.

\subsection{The case $N=3$}

At $N=3$, ${\mathcal X}(t)=t^3+e_2t-e_3$ with
\begin{equation}
e_2=-\tfrac12\tr(\phi^1)^2,\qquad e_3=\det\phi^1=\tfrac13\tr(\phi^1)^3 .
\end{equation}
Choosing the pair is the same as choosing the spectator. If the pair is $\{a,b\}$ and the spectator is $c$, then $\lambda_a+\lambda_b=-\lambda_c$ and $\lambda_a\lambda_b=\lambda_c^2+e_2$, so all pair data depend on a single variable $x=\lambda_c$:
\begin{equation}
r(t)=t-x,\qquad \Delta={\mathcal X}'(x)=3x^2+e_2,\qquad \tilde v(t)=-(t+2x),\qquad D(Y)=Y,
\end{equation}
\begin{equation}
\mathcal P=2x^2-x\phi^1-(\phi^1)^2,\qquad\mathcal P'=(\phi^1)^2+x\phi^1+x^2+e_2 .
\end{equation}
Here $\mathcal P/\Delta$ is the projector onto the selected pair and $\mathcal P'/\Delta$ is the rank-one projector onto the spectator. The scaled letters \eqref{eq:scaledletters} are
\begin{equation}
\begin{aligned}
L^m&=\mathcal P\phi^m,\qquad F_2=\mathcal P\psi_2,\qquad F_3=\mathcal P\psi_3,\\
F_1&=\mathcal P\Big[(3x^2+e_2)^2\psi_1+(\phi^1+2x)\big(\psi_3\,\mathcal P'\phi^3+\psi_2\,\mathcal P'\phi^2\big)\Big].
\end{aligned}
\end{equation}
Writing
\begin{equation}
\Xi^m_k=\tr(F_kL^m)-\tfrac12\tr(F_k)\tr(L^m),\qquad \Theta_{ba}=\tr(F_bF_aF_a),
\end{equation}
and reading off $n_1$ for each monomial of \eqref{eq:O24trace}, the polynomial $H(x)\equiv\Delta^{13}O_S$ is
\begin{equation}
H(x)=-16i\Big[\big(\Xi^1_1-\Delta^2\Xi^2_2\big)\Xi^1_3\,\Theta_{21}+\Delta^4\big(\Xi^2_2-\Xi^3_3\big)\Xi^2_1\,\Theta_{32}+\big(\Delta^4\Xi^3_3-\Delta^2\Xi^1_1\big)\Xi^3_2\,\Theta_{13}\Big],\label{eq:HN3}
\end{equation}
with $\Delta=3x^2+e_2$. Reduce $H$ modulo ${\mathcal X}$, using $x^3=-e_2x+e_3$:
\begin{equation}
H(x)\equiv h_0+h_1x+h_2x^2\pmod{{\mathcal X}(x)} .
\end{equation}
The coefficients $h_0$, $h_1$ and $h_2$ are gauge-invariant multi-trace polynomials in the BMN letters, with coefficients polynomial in $e_2$ and $e_3$. They do not depend on $K$. Since $\sum_c f(\lambda_c)=\tr f(\phi^1)$ for any polynomial $f$ with gauge-invariant coefficients, the sum over the three spectators gives
\begin{equation}
O_3^{(K)}=\sum_c\mathcal X'(\lambda_c)^{K-13}H(\lambda_c)
=h_0\tr M+h_1\tr(\phi^1M)+h_2\tr\big((\phi^1)^2M\big),\qquad
M=\big(3(\phi^1)^2+e_2\big)^{K-13},\label{eq:O3final}
\end{equation}
for every $K\geq13$. For $K=13$ this reduces to $O_3^{(13)}=3h_0-2e_2h_2$. Equation \eqref{eq:O3final} is the $N=3$ case of \eqref{eq:doubletrace}, written in terms of the single spectator variable. Together with \eqref{eq:HN3} and the definitions above, it specifies $O_3^{(K)}$ completely as a polynomial in the BMN letters.

\end{document}